# Uncertainty Quantification in Atomistic Modeling of Metals and its Effect on Mesoscale and Continuum Modeling – A Review

Joshua J. Gabriel[1*], Noah H. Paulson[1], Thien C. Duong[2], Francesca Tavazza[3], Chandler A. Becker[4], Santanu Chaudhuri[5,6], and Marius Stan[1]

[1]*Applied Materials Division, Argonne National Laboratory, Lemont, IL – 60439, USA*
[2]*Energy and Global Security, Argonne National Laboratory, Lemont, IL – 60439, USA*
[3]*Material Measurement Laboratory, National Institute of Standards and Technology, Gaithersburg, MD – 20899, USA*
[4]*Office of Data and Informatics, Material Measurement Laboratory, National Institute of Standards and Technology, Gaithersburg, MD – 20899, USA*
[5] *Manufacturing Science and Engineering, Energy and Global Security, Argonne National Laboratory, Lemont, IL – 60439, USA*
[6] *Civil, Materials, and Environmental Engineering, University of Illinois at Chicago, Chicago, IL – 60607, USA*
*Corresponding author: jgabriel@anl.gov

**Abstract**
The design of next-generation alloys through the Integrated Computational Materials Engineering (ICME) approach relies on multi-scale computer simulations to provide thermodynamic properties when experiments are difficult to conduct. Atomistic methods such as Density Functional Theory (DFT) and Molecular Dynamics (MD) have been successful in predicting properties of never before studied compounds or phases. However, uncertainty quantification (UQ) of DFT and MD results is rarely reported due to computational and UQ methodology challenges. Over the past decade, studies have emerged that mitigate this gap. These advances are reviewed in the context of thermodynamic modeling and information exchange with mesoscale methods such as Phase Field Method (PFM) and Calculation of Phase Diagrams (CALPHAD). The importance of UQ is illustrated using properties of metals, with aluminum as an example, and highlighting deterministic, frequentist and Bayesian methodologies. Challenges facing routine uncertainty quantification and an outlook on addressing them are also presented.

1. Introduction

Integrated Computational Materials Engineering (ICME) describes the design of materials for target properties by the coupled use of experiments, computational simulations and data driven techniques. Atomistic simulation workflows that cross multiple time- and length-scales are becoming popular for the determination of physical properties critical to ICME. One often overlooked tenet of ICME, however, is the reliable quantification of uncertainties of material properties. This is especially important for the design of metals that are used in transportation, structural, health, and energy industries due to the mission critical nature of the materials performance and the potential for loss of life should failures occur [1, 2, 3]. In this review, we first introduce the terminology used to express uncertainties in the atomistic simulations (Density Functional Theory (DFT) and Molecular Dynamics (MD)) literature. Next, we discuss the flow of information between atomistic simulation methods and mesoscale (Phase Field Modeling (PFM)) and thermodynamic (Calculation of Phase Diagrams (CALPHAD)) models, in the context of calculated thermodynamic properties. We then introduce the uncertainty quantification approaches, both Bayesian and frequentist, that have been made in the context of PFM and CALPHAD. In subsequent sections, we describe the uncertainty quantification approaches in DFT (Section 2) and MD (Section 3), and examine the uncertainties reported for the thermodynamic properties of aluminum with atomistic simulation methods. We then describe how atomistic simulation data with uncertainties have been used in CALPHAD (Section 4) and PFM (Section 5). Finally, in Section 6, we conclude by discussing the challenges with regards to the regular use of uncertainty quantified data in developing thermodynamic models with these methods and present our outlook on how some of these challenges can be addressed.

1.1 Types of Uncertainty

Formal approaches to the quantification of uncertainty continues to be an active area of development for atomistic simulations [4, 5]. Inherent to these activities is the definition of the types and sources of uncertainties/errors. A comprehensive review of the uncertainty concept in the context of multiscale materials simulations, their types and sources are beyond the scope of this article. For more information on these topics, the reader is encouraged to access the excellent book by Wang and McDowell [6]. For the purpose of this review, we adopt the broad classification of uncertainties as epistemic and aleatoric uncertainties. In addition, a large variety of terms are employed in the literature for uncertainty quantification in atomistic simulations. These include *systematic error, random error, precision, accuracy, convergence error, numerical precision, controlled and uncontrolled approximations, model uncertainty* and *parametric uncertainty*. It is not always clear how these terms are related to the broad classification into epistemic and aleatoric uncertainty and so we introduce them as follows:

a. Epistemic uncertainties are uncertainties caused by a lack of knowledge stemming from data and/or model form insufficiencies and the subjectivity of model parameter choice due to experience. Data and model form insufficiencies are caused by computational cost considerations for data acquisition or model evaluation, or a combination of both. The uncertainties caused by computational cost considerations are controllable and hence are also referred as *controlled approximations* in the computational modeling literature. The bias in the model is referred as the *model form uncertainty* and it manifests itself as a *systematic error*. The error itself is expressed as the *accuracy* if the ground truth is known. If the error originates as a result of computational cost considerations, this error manifests as the *convergence error* or the *numerical precision error*. Sometimes however, the epistemic uncertainty cannot be reduced predictably, and such errors are referred as *uncontrolled approximations*. The subjectivity of model parameter choice is an example of *parametric uncertainty* which is epistemic in nature.

b. Aleatoric uncertainty is *random error* that can be quantified in the form of probability distributions. They are caused by stochastic aspects of a computational experiment or set up of a model. Variability in the structure of a material with defects is an example of stochastic aspects of a computational experiment. The aleatoric component of *parametric uncertainty* is related to the distribution of a model's parameters that best match the data.

Uncertainty quantification in DFT has historically dealt with epistemic uncertainty quantification using descriptive statistics, though probabilistic uncertainty quantification approaches for inferential statistics continue to be developed. Uncertainty quantification in MD, CALPHAD and PFM on the other hand, has also dealt with probabilistic uncertainty quantification for inferential statistics. Frequentist and Bayesian statistics are the two dominant approaches to probabilistic uncertainty quantification for inferential statistics. Frequentist statistics works under the assumption that a given model is deterministic (or that certain parameters have defined probability distributions), and that through large numbers of observations the probability of the data being supported by the model can be found, or an interval in which the true model parameters reside can be identified with a certain probability. In contrast, Bayesian statistics assumes models to be probabilistic, and uses observed data to update prior beliefs about the probability distribution of model parameters and other quantities. CALPHAD in particular makes use of both frequentist [7] and Bayesian [8] approaches to uncertainty quantification.

Bayesian statistics and Bayesian concepts are highlighted in the remainder of this review. Consequently, a brief description of Bayes' Theorem is provided below. For a model M parameterized by $\vec{\theta}$, Bayes' Theorem,

$$\Pr(\vec{\theta}|\vec{D}, M) = \frac{\Pr(\vec{D}|\vec{\theta}, M) \Pr(\vec{\theta}|M)}{\Pr(\vec{D}|M)} \quad (1)$$

describes the posterior probability distribution $\Pr(\vec{\theta}|\vec{D}, M)$ of the model parameters given the observation of data $\vec{D}$, where $\Pr(\vec{D}|\vec{\theta}, M)$ is the likelihood of the data given a specific parameter set, $\Pr(\vec{\theta}|M)$ is the prior assumed distribution of the parameters before the observation of data, and $\Pr(\vec{D}|M)$ is the marginal likelihood, calculated by integrating the numerator of the expression across the entire parameter space. Given certain choices of the model form, likelihood, and prior distributions it is possible to derive an analytical expression for the posterior distribution, , but in the majority of cases the posterior must be evaluated through numerical means, most typically Markov chain Monte Carlo (MCMC). The critical choices that affect the posterior and therefore predictive uncertainties are those of the model form, the prior distributions, and the likelihood function. Of these, only the likelihood considers the data and therefore will be of most interest in understanding the connections between CALPHAD and DFT uncertainty. It is common practice to assume a Gaussian likelihood function (although the Student's t-distribution may be used to increase robustness to outliers), and therefore the variance must be specified [9]. As the likelihood represents the distribution of the data around the mean model, this variance is equivalent to the uncertainty in the data. Two common choices are to fit a variance hyperparameter in the Bayesian inference, or to simply use the reported errors as an estimate.

**1.2 Uncertainty propagation between interdependent simulation methods**

Uncertainty propagation between the individual components of multi-scale simulations of materials structure is important because of the sensitivity of phase stability models to errors as small as 1 meV/atom, which is the resolution of energy accuracy required to determine phase transitions [10]. Typically, multi-scale atomistic simulations are viewed as traversing increasing length and time scales along a straight line as shown in Figure 1 [11], with higher length and time scale simulations depending on lower length and time scale simulations. However, in practice, information can be passed between the methods from a higher scale method to a lower scale method or by skipping a length or time scale in between. Hence, for the purposes of uncertainty quantification and propagation between the four methods of DFT, MD, CALPHAD and PFM, we propose viewing the methods as four interconnected points of a rectangle, as shown in Figure 2. DFT, as the name suggests, calculates properties based on functionals of electron density [12]. In contrast, MD simulations use Newton's classical equations of motion, with an interatomic potential that models the interactions between atoms, at specified conditions, such as temperature and pressure [13, 14, 15, 16]. CALPHAD describes the use of Gibbs energy models for phases of interest as a function of composition, temperature, and pressure to predict the stability and thermodynamic properties of pure components and mixtures through coupled equilibrium calculations. The phase field method (PFM) is used to model the evolution of microstructures [17]. The following properties of metals are exchanged between these methods: heat capacity at constant pressure ($C_p$), enthalpy ($H$), free energy (Gibbs, $G$ and Helmholtz, $F$), phase transition temperatures such as the melting point ($T_m$), diffusion coefficients ($D$), interfacial energies ($\gamma$), and elastic constants ($C_{ij}$). As shown in Figure 2, each of these properties can be determined independently by each of DFT and MD, or by a combination of methods. For example, DFT can be used to parametrize interatomic potentials for molecular dynamics by calculating properties such as an equation of state, or the energy, forces, and stresses that describe the potential energy surface. In turn, the interatomic potential can be used to calculate enthalpies ($H$), Helmholtz free energies ($F$), diffusion coefficients ($D$) and interfacial energies ($\gamma$), at a given temperature and pressure by propagating a material system over a long enough time scale. These properties can also parametrize a PFM to describe the evolution of microstructures. Some of these thermodynamic properties ($C_p$, $T_m$, $F$) can also be calculated directly with DFT based molecular dynamics or the quasi-harmonic approach, but system size and calculation time remains a challenge for average computational budgets.

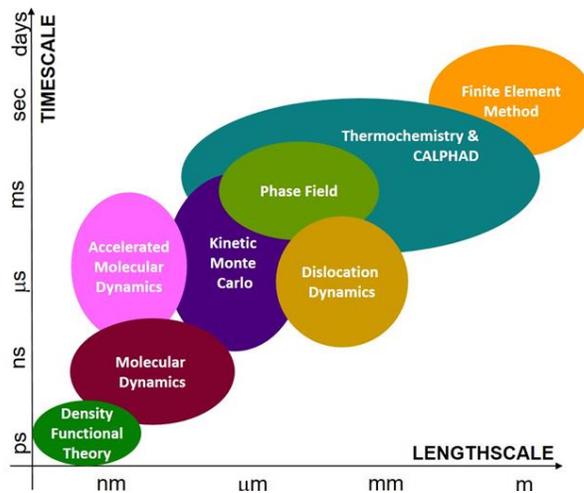

*Fig. 1 Atomistic simulation methods and the spatio-temporal regions of interest covered by each atomistic simulation method. Reprinted with permission from [11] .*

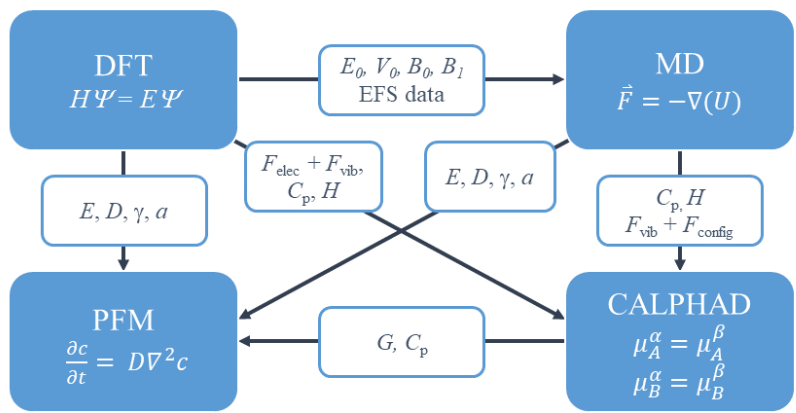

*Fig. 2 Properties passed between simulation methods show the uncertainty transferred forward and inversely between methods. E is the total energy from DFT, D is the diffusion constant, F is the Helmholtz free energy, G is the Gibbs free energy, either of which can have electronic (elec), vibrational (vib) and configurational (config) components, $C_p$ is the heat capacity at constant pressure, H is the enthalpy, γ is the interfacial or surface energy, a is the lattice parameter and μ is the chemical potential. "EFS" corresponds to energy, force and stress data calculated with DFT and used to fit interatomic potentials in MD and $E_0$, $V_0$, $B_0$, and $B_1$, are respectively the cohesive energy, equilibrium volume, bulk modulus and the pressure derivative of the bulk modulus that are considered as fitting targets for MD.*

## 2. Uncertainty Quantification in DFT and impacts on MD, CALPHAD, and PFM

DFT calculates materials properties by solving for the electronic ground state of the material. To this end, important approximations are made to describe the number of electrons and the interactions between them. In this section, we first describe how uncertainties have been quantified for DFT computed properties in the context of choices to these approximations. We then describe the uncertainties for properties that are used in MD, CALPHAD, and PFM.

### 2.1 Uncertainty Quantification Approaches for DFT computed properties

Uncertainty quantification approaches for DFT computed properties have focused largely on quantifying epistemic uncertainties. These epistemic uncertainties are caused by the choice of exchange correlation functional, pseudopotential or all-electron treatment of the interacting electrons, and by the choice of calculation convergence parameters. The first two choices define the physics of the system and hence result in model form errors. The choice of calculation convergence parameters results in numerical precision errors. To a lesser extent, there have been attempts at quantifying aleatoric uncertainties arising from variability in the representation of the simulation box describing the system under study. In this section, we review how errors due to these choices have been quantified.

Modelling interacting electrons is beyond current capabilities, and so, in DFT, electrons are approximated by an auxiliary system of noninteracting electrons, where each of them immerses in an effective single-particle potential. Such a potential contains an electron–electron Coulomb repulsion term and an exchange–correlation potential term that approximate all the many-particle interactions. Many new exchange correlation functionals [18, 19, 20] have been developed in recent years to improve property predictions especially for certain elements. Whenever a new exchange correlation functional or implementation thereof is introduced, benchmark studies are performed by comparing the new exchange correlation functional to existing ones and to experimental data. In these benchmark studies, the uncertainties are quantified with respect to a chosen gold standard, using statistical quantities such as the Mean Absolute Error (MAE) and Mean Absolute Percentage Error (MAPE). These are epistemic uncertainties contributing to the model form errors. For a given material and choice of exchange-correlation, this uncertainty is fixed; it can be reduced only by selecting a better exchange-correlation functional.  For example, Tran *et al.* [21] compared lattice constants, bulk moduli and cohesive energies for 63 new and old functionals from different classes:

the Local Density Approximation (LDA), Generalized Gradient Approximation (like Perdew Burke Ernzerhof, called GGA-PBE), meta-GGAs (like the Strongly Constrained and Appropriately Normed (SCAN) and meta Bayesian Error Estimation Functionals (BEEF)) and hybrid functionals (like PBE0), with and without dispersion corrections (Van der Waals (vdW) corrections). They found that for strongly bound solids, GGA is as accurate as higher-level functionals, while meta-GGA functionals are needed for finite systems, and dispersion-corrected ones are necessary for an accurate description of weakly bound materials. Janthon *et al.* [22] determined that meta-GGA and meta-NGA (Nonseparable Gradient Approximation) functionals provide good description of molecular crystals while also having accuracy comparable to the GGA functional for transition metals. Additionally, the Bayesian Error Estimation Functional [20] provides an ensemble averaged error estimate for property prediction, which is an intrinsic uncertainty independent of an experimental reference. The disagreement between the DFT prediction for a specific functional and experimental data can be exploited for materials design as well. For example, in Choudhary *et al.*, the disagreement in lattice constants predictions was used as a screening criterion to identify exfoliable materials [23].

Although DFT avoids dealing with the many-body problem, solving the Kohn-Sham (KS) equations for all the electrons in the system is very computationally intensive. Therefore, in addition to the all-electron approach, where all the electrons in the system are taken into account, a second approach is commonly used: the pseudopotential approach [24, 25, 26]. Here the (KS) equations are only solved for the valence electrons, while the non-valence electrons are treated as a frozen core. Like exchange correlation functionals, pseudopotential and all electron approaches continue to be developed. A large study [27] focused on accuracy across different DFT codes showed that, for the same exchange-correlation functional (PBE), predictions from recent codes agree very well with each other, provided that the most recent version of the proper pseudopotential is used. Specifically, pairwise differences in equation of state (EOS) between codes are comparable or smaller than those between high-precision experiments. If older versions of pseudopotentials are used, differences between codes become substantially larger. This work introduced a quality metric for the comparison of different DFT codes, known as the delta gauge, that continues to serve as a quality metric for newly developed DFT codes.

In addition to pseudopotentials and exchange correlation functionals, other parameters are key in determining DFT numerical precision and accuracy. Examples of these parameters are the density of the *k*-point mesh to perform the energy integration and the number of plane waves used to expand the wave function in plane waves codes. Most DFT databases determine these parameters for a few key materials and then use them for all compounds in the repository. One noticeable exception is the JARVIS-DFT database [28], where such parameters are converged for each included material. Typically, choices are made based on the requirement of achieving an energy convergence of 1 meV/atom, which is the energy difference over which phase transitions take place. Gabriel *et al.* [29], however, showed that a k-point density choice sufficient for the convergence of energy does not always guarantee convergence for a derived property of interest. For example, the pressure derivative of the bulk modulus is converged to 1 % only when the energy is converged to less than 1 meV/atom. This work showed that the precision of the equilibrium volume, bulk modulus, and the pressure derivative of the bulk modulus correlate comparably well with the k-point density and the precision of the energy, following an approximate power law. They also established that common k-point density choices in high throughput DFT databases result in precision for the volume of 0.1 %, the bulk modulus of 1 %, and the pressure derivative of 10 %.

Fewer studies have attempted quantifying aleatoric uncertainty in DFT calculations. One approach to aleatoric uncertainty was undertaken to capture the effect of variability in the arrangement of atoms in amorphous materials [30]. The aleatoric uncertainty was found to depend on the system size and could be as much as a factor of three larger than epistemic uncertainties for small systems.

**2.2 Reported Uncertainties for DFT calculated properties relevant to MD, CALPHAD and PFM**

In this section, we describe the uncertainties of properties computed with DFT and that are used in MD, CALPHAD and PFM. We begin with properties that serve as targets for the interatomic potential used in MD. Next, we describe properties that are used in CALPHAD and PFM.

For MD, the interatomic potential describes the interactions between atoms. Classical interatomic potentials, such as the Embedded Atom Method (EAM), a popular interatomic potential for metals, are fit to reproduce target properties from experiment and DFT calculations. Among these target properties are the equation of state properties namely, cohesive energy ($E_0$), equilibrium volume ($V_0$), bulk modulus (B), the pressure derivative of the bulk modulus ($B_1$), and the elastic constants ($C_{ij}$). The pseudopotential approach and the generalized gradient approximation functional of Perdew-Burke-Ernzerhof (PBE) [31] is the most widely used functional in DFT materials data repositories [32, 33, 34, 35]. Using this choice of exchange-correlation, Lejaeghere *et al.* [36] estimated errors on equation of state properties and key elastic constants ($C_{11}$, $C_{12}$, $C_{33}$, $C_{13}$, and $C_{44}$), and performed linear regressions through least-square fits between experiments and calculated properties for elemental crystals in their stress-free ground state. The experimentally measured properties were first extrapolated to 0 K and corrected for zero-point vibrations. From the fits, the slope and the scatter with respect to the regression line were determined. The difference between the slope and the value "*1*" gave the systematic error, while the standard deviation of the scatter gave the residual error bar. The main source of this scatter is the model form error due to the choice of the exchange correlation functional, which, for a given choice, performs best for certain elements and worse for others. The systematic error is the result of a specific implementation of the DFT code, which is reflected in the choice of algorithm to solve the Kohn-Sham equations, the chosen pseudopotential, etc. By grouping the elemental crystals into eight classes based on common physical properties, Lejaeghere *et al.* determined what structure types are well described by DFT calculations using PBE and excluded the others (strongly correlated materials and materials where dispersion interactions are essential, i.e. ionic crystals and noble gases) from their analysis. Magnetic materials were not excluded from the analysis but do show larger scatter than other groups with respect to cohesive energy, pointing out that PBE, and possibly other current Generalized Gradient Approximation (GGA) functionals, are not able to describe magnetic compounds as well as other types of materials. The largest relative systematic deviation from slope=1 was found for the bulk modulus (−4.9 %) and for its pressure derivative, $B_1$ (+4.8 %), followed by the equilibrium volume (+3.6 %) and $C_{ij}$ (−2.0 %), where $C_{ij}$ is the mean over the key elastic constants. The slope was found to be 1 for cohesive energy. A positive (negative) sign means that PBE tends to overestimate (underestimate) the quantity. Lastly, the authors noted that elements with the highest deviation in cohesive energy did not always show highest deviations in the other examined properties. A similar study related experimentally measured melting points to DFT calculated cohesive energies aiming to develop a semi-empirical model that could predict experimental melting points from DFT calculated cohesive energies. Prediction errors can be as small as 10 K for some metals and as large as 750 K for other metals [37].

For CALPHAD, the heat capacity (Cp) can be obtained from the quasi-harmonic approximation to the free energy. The heat capacity at constant pressure can be computed from the free energy and can more easily be compared to experiment. In a recent study, the heat capacity was determined using the Bayesian error estimation functional and the quasi-harmonic approach to aluminum [38]. Although computationally expensive for regular practice, melting points and phase transition temperatures with uncertainty have been determined from the trajectories of *ab initio* molecular dynamics (AIMD) [39]. In Table 1, we tabulate a subset of predicted thermodynamic properties of aluminum such as the melting point, heat capacity and their reported errors from DFT and MD. We intend Table 1 as examples of reported uncertainties in the DFT and MD literature on the enthalpy and heat capacity of aluminum, but by no means an exhaustive collection of all studies. For DFT, major approximations are the different choices of the pseudopotential (PP), exchange correlation functional (XC), and basis set, and *k*-points density expressed as choice of Monkhorst-Pack (MP) [40] mesh. For MD, only one interatomic potential is mentioned as an example comparison with DFT, though we note that a number of interatomic potentials exist for aluminum, most of which have not been evaluated systematically for these properties with reported uncertainties [41].

For PFM, the interfacial energy, lattice parameter, elastic tensor, chemical potential, and diffusion coefficient can be derived from DFT calculations. Diffusion coefficients can be calculated from density functional theory metadynamics [35] and DFT powered MD simulations [36]. Comparing the surface energies for elemental crystals, Tran *et al.* [42] created a database of Wulff crystal shapes of the elements and found that the maximum convergence error with respect to DFT calculation inputs, under the widely used GGA-PBE, was 0.02 J/m$^2$.

*Table 1 Examples of properties and uncertainties reported for select papers for thermodynamic properties of aluminum from DFT and MD, showing some of the types of uncertainties and approximations that are helpful to report. Reported enthalpies and estimated heat capacities from [43] [44] an Embedded Atom Model (EAM) interatomic potential of aluminum from MD simulations. The heat capacity was estimated as the derivative of the enthalpy with respect to temperature.*

| Publication Year | Property | Estimate (Experiment) | Major Approximations |
|---|---|---|---|
| 1998 [39] | Melting temperature | 890 ± 20 K (933.7) | PP: PAW<br>XC: LDA<br>Basis set: planewave,<br>$k$-points: MP 6 x 6 x 6 |
| 2008 [45] | Heat Capacity, Cp at 300 K | 23.86 J mol$^{-1}$K$^{-1}$ | PP: PAW<br>XC: LDA<br>Basis set: planewave,<br>$k$-points: MP 57 x 57 x 57 |
| 2008 [45] | Heat Capacity, Cp at 300 K | 24.36 J mol$^{-1}$K$^{-1}$ | PP: PAW<br>XC: GGA-PBE<br>Basis set: planewave,<br>$k$-points: MP 57 x 57 x 57 |
| 2019 [38] | Heat Capacity, Cp at 300 K | 24.10 ± 1.04 J mol$^{-1}$K$^{-1}$ | PP: PAW<br>XC: BEEF-vdW<br>Basis set: planewave<br>$k$-points: 30 Å$^{-1}$ |
| 2019 [38] | Heat Capacity, Cp at 900 K | 29.03 ± 5.23 J mol$^{-1}$K$^{-1}$ | PP: PAW<br>XC: BEEF-vdW<br>Basis set: planewave<br>$k$-points: 30 Å$^{-1}$ |
| 2010 [43] [44] | Enthalpy at 900 K | 28.3 ± 0.3 kJ mol$^{-1}$ | EAM potential 'Al1' from [66]; crystalline, amorphous structural properties. Liquid structure factors from XRD |
| 2010 [43] [44] | Heat Capacity $C_p$ at 900 K | 32.96 ± 0.4 J mol$^{-1}$K$^{-1}$ | EAM Al1 |
| 2010 [43] [44] | Enthalpy at 1000 K | 31.4 ± 0.3 kJ mol$^{-1}$ | EAM Al1 |
| 2010 [43] [44] | Heat Capacity $C_p$ at 1000 K | 32.49 ± 0.5 J mol$^{-1}$K$^{-1}$ | EAM Al1 |

## 3. Uncertainty in Molecular Dynamics and Impacts on CALPHAD and PFM

In an MD simulation, the interatomic potential (IP) function defines the interactions between atoms [46]. The gradient of the IP ($\vec{F} = -\nabla(U)$) determines the velocity of atoms and how the thermodynamic state of a system of atoms occupying a volume $V$ evolves with time $t$ to a state defined by the total energy E(V,t,p,T), where $T$ is temperature and $p$ is pressure. In an MD simulation, choices are made for the interatomic potential, the pathway to the desired thermodynamic state, the equilibration time to get to that state, and the boundary conditions of the simulation itself. Each of these choices cause both epistemic and aleatoric uncertainties. In this section, we review first the approaches to quantify these uncertainties. Then, we discuss uncertainties for thermodynamic properties and describe their impact on CALPHAD and PFM.

### 3.1 Uncertainty Quantification Approaches for MD simulations

Uncertainty quantification approaches for MD have largely focused on the choice of interatomic potential, the parameters that parametrize each potential, and descriptive statistical measures of the outputs of a simulation. The choice of interatomic potential defines the underlying physics and is hence a model form uncertainty. Sensitivity analysis approaches have also been applied to the parameters that define the interatomic potentials. These studies pertain to quantifying the parametric uncertainty with Bayesian statistics. For the purpose of this review, we classify MD simulations into three categories: classical MD simulations, machine learning force field molecular dynamics (MLFF-MD), and reactive molecular dynamics (RMD). In this subsection, we review model form and parametric uncertainty quantification approaches in the context of classical MD, MLFF-MD and RMD.

Uncertainties in classical MD primarily occur for the following reasons: a) the choice of the interatomic potential for a given MD simulation; b) the choice of inputs outlined in Section 1 for DFT calculations of reference properties and the experimental data, which was used to fit the interatomic potential; (c) simplifications to the modeled material when compared to the experimentally characterized material; d) differences in the testing procedures between experiments and simulations; and e) data analysis technique [47]. Most studies show that the choice of the force field is the main factor that affects the predictions of material properties. [48].

Interatomic potentials are derived to target certain experimental or DFT calculated properties for a limited number (a calibration data set) of known crystal structures and defects. As such, their transference to structures or property predictions outside the calibration data set can be questionable [49]. In addition, there is uncertainty in measurement and/or the DFT calculated data, as well as the assumed functional form of the interatomic potential. The first IPs (pertaining to classical MD) were fit to simple functional forms of interatomic distances and/or bond angles to reproduce experimental data; new potentials (pertaining to MLFF-MD) are fit to DFT data such as atomic forces, energies and stresses, sometimes using flexible functional forms or complex descriptions of local atomic environments. Some of these flexible functional forms such as Gaussians [50] yield intrinsic uncertainties to the predicted energy and forces, which further guide the selection of calibration data [51].

The quantification of parametric uncertainty for single potentials has been undertaken in several cases [52, 53] while Bayesian frameworks have also been proposed for a variety of interatomic models and force fields [54, 55, 56]. Furthermore, quantification of uncertainty due to the potential fitting reference set [57] was augmented by propagation of parametric uncertainties to MD outputs [58]. Recent efforts have focused on fitting interatomic potentials to data and subsequently quantifying the uncertainty [59]. These efforts contributed to the uncertainty quantification and potential development by providing an open source implementation of the framework proposed by Frederiksen *et al* [54]. The uncertainty in the MD model parameters propagates to predictions of properties such as density, thermal expansion coefficient, isothermal compressibility, enthalpy and viscosity. The level of uncertainties in relation to the uncertainties observed in the experimental quantities is partly due to the large fluctuation of these properties arising from short time intervals used in MD simulations [58]. Frederiksen *et al.* applied concepts from Bayesian statistics to estimate error bars on properties predicted through MD. They compared three different potentials and assigned independent normal likelihood to the model discrepancies from DFT or experiment values [54].

A good measure of the confidence in the model predictions consists of evaluating the uncertainty in the effective potential. Longbottom, *et al.*, have demonstrated this technique using three potentials for nickel: two simple pair potentials, Lennard-Jones and Morse, and a local density dependent embedded atom method potential [59]. They were successful in developing a potential ensemble fit to DFT calibration data to calculate the uncertainties in lattice constants, elastic constants and thermal expansion of nickel. A different approach was used by Reeve, *et al.*, who used functional derivatives to quantify how thermodynamic outputs of a MD simulation depend on the potential used to compute atomic interactions [60]. In this approach, the sensitivity of the quantities of interest (QOI) is evaluated with respect to the input functions as opposed to its parameters, as done with traditional uncertainty quantification methods. Reeve, *et al.*, were successful in demonstrating the power of this approach under three different thermodynamic conditions: a crystal at room temperature, a liquid at ambient pressure, and a high-pressure liquid.

Rizzi, *et al.*, [56] focused on the forward propagation of MD uncertainty starting with quantifying the effect of intrinsic (thermal) noise and the parametric uncertainty in MD simulations. The parametric uncertainty was assumed to originate from IP parameters as standard uniform random variables. The thermal fluctuations inherent in MD simulations, combined with parametric uncertainty, resulted in noisy MD predictions of bulk properties. In subsequent studies, Rizzi, *et al.*, [61] explored the inference of small-scale, atomistic parameters, based on the specification of large, or macroscale, observables. The results demonstrated that a suitable choice of the observables allows the recovery of "true" parameters with high accuracy even with low-order surrogate models. MD evolution equations are non-linear and strongly [62] coupled, as discussed by Grogan, *et al.* [63]. In their study, they made detailed numerical comparisons between full classical MD simulations and MD simulations using large-scale approximations. The reliability of these methods was evaluated by measuring the differences between full, classical MD simulations and those based on these large-scale approximations. The study demonstrated the existence of computationally efficient large-scale MD approximations that accurately model certain large-scale properties of the molecules such as energy and linear and angular momenta.

Stochastic methods are also used to evaluate uncertainty of MD simulations. For example, a methodology enabling the robust treatment of model form uncertainties in MD simulations was proposed by Wang *et al*. [62]. The approach consists of properly randomizing a reduced-order basis, obtained by the method of snapshots in the configuration space. A multi-step strategy to identify the hyperparameters in the stochastic reduced-order basis was further introduced, enabling the robust, simultaneous treatment of parametric uncertainties on a set of potentials [62] Furthermore, uncertainty quantification in non-equilibrium phenomena, such as thermal transport, was studied to estimate bulk thermal conductivity via non-equilibrium molecular dynamics (NEMD) [52].

Reactive molecular dynamics (RMD) simulations can also be subject to multiple sources of error and the approach in tracking UQ is somewhat more involved in comparison to other classical MD simulations. Many reaction networks can progress in multiple different pathways leading to entirely different products and product distributions at the end of RMD trajectories. Multi-objective optimization of force field parameters and uncertainty quantification can be merged to provide a standardized UQ capability for reactive simulations [55]. In case of extremely fast reactions of thermal deflagration, the velocity of propagation can make a significant difference unless the time steps for RMD simulations are restricted to 0.1 fs and below for obtaining consistent results [64, 65, 66]. Subsequently, mirrored atomistic RMD and continuum simulations show that average of rates, temperature, and pressure can also carry significant differences due to atomistic scale fluctuations in averages calculated using a control volume (CV) approach and propagated to the continuum scale [65, 67, 68]. Integration schemes and polynomial fitting of rates of reactions are prone to their own numerical error. However, it is important to develop UQ approaches for MD and RMD simulations to develop better methods for taking averages from a stochastic and fluctuating domain in an atomistic ensemble simulation, and upscaling them for use in continuum scale.

**3.2 Reported MD Uncertainties and their impact on PFM and CALPHAD**

For CALPHAD, as shown in Figure 2, the heat capacity (Cp), enthalpy and free energy can be estimated with MD simulations. The enthalpy and free energy are obtained as direct outputs of an MD simulation. The heat capacity can be obtained as the derivative of the enthalpy from MD runs performed at different temperatures. In Table 1, we tabulate examples of uncertainties reported from statistical averaging of the enthalpy of aluminum. Such an approach, though simple, we note is not yet widely reported for different interatomic potentials for these properties. The magnitude of the error bars is dependent on the equilibration time of MD simulation runs, which is another input parameter to the MD simulation.

The EAM potential is one of the favorite choices for MD modeling of metals. Dhaliwal *et al.* [69] have performed uncertainty and sensitivity analysis of mechanical and thermal properties computed through EAM. They concluded that the predictions can be sensitive to the small perturbations in IP parameters. In order to make MD predictions for complex material systems more reliable, they studied in detail the variations in the experimental values of various mechanical and thermal properties of FCC Al. The probability distributions of the IP parameters were obtained using a Bayesian statistical framework and the reliability of potential parameters was assessed by performing MD simulations for a range of mechanical and thermal properties, using perturbed potential parameters. A comparison of the computed properties to experimental and first-principles data revealed that higher order properties such as grain boundary formation energy are sensitive (with variance of the order $10^5$) to 1% perturbations. It was also observed that the QOI computed through EAM was highly sensitive to changes in the IP parameters. For example, perturbing the IP parameters by 1 % resulted in grain boundary formation energy variations as high as 85 % of the original fitted values. Tran *et al*. used the interval-based approach for uncertainty analysis in EAM potentials [70]. The uncertainty in tabulated EAM potential was captured by analytical forms of error generating functions and the method was applied to aluminum, resulting in accurate stress–strain curves.

MD simulations have been coupled with PFM to describe the evolution of microstructures. As shown by Zhang *et al.*, in a study of solidification dynamics of Cobalt using EAM potential, the microstructures can be slightly different for different choices of MD simulation inputs such as the time step, thermostat parameters, and domain decomposition scheme for the atoms. Differences in these inputs, under the same cooling rates, was shown to yield nanocrystalline, lamellar, or microcrystalline grain structures, due to small differences in nucleation location and growth possible under severely undercooled regions [71]. Hence, extensive care is needed to manage the uncertainties by controlling time steps, thermostat parameters, and even domain decomposition schemes before a converged observation of microstructure with same potential energies is achieved. These differences become more pronounced when microstructure evolution is modelled in additive manufacturing of Co-alloys such as AF75 alloys (Co-Cr-Mo) using phase field methods. The robustness of PFM predictions is affected by model form and parametric uncertainties. Tran *et al*. have studied and quantified the uncertainty of PFM predictions of Al-Cu microstructure evolution [72]. A surrogate model was used to interpolate QOIs such as perimeter, area, primary arm length, and solute segregation, as functions of thermodynamic and process parameters. The effect of parametric uncertainty on the Al–Cu dendritic growth during solidification simulation was investigated. The results showed that the dendritic morphology varies significantly with respect to the interface mobility and the initial temperature.

**4. Uncertainty quantification and Bayesian assessment of atomistic data for CALPHAD**

CALPHAD serves a critical role in the design and improvement of engineering alloys, and as an input to other simulation approaches (e.g., precipitation simulations and the phase field method). In this method, thermodynamic equilibrium is given by Gibbs' rules. For a binary system with components A and B and phases α and β, the required equality of chemical potentials μ is given by: $\mu_A^\alpha = \mu_A^\beta$ and $\mu_B^\alpha = \mu_B^\beta$. CALPHAD models are calibrated with two classes of information, i) phase stability/transition measurements, and ii) the thermodynamic properties of phases and mixtures. It is this second category of information that is most useful for the extrapolation to metastable regimes and multicomponent systems, and simultaneously the most difficult to access experimentally. For this reason, CALPHAD practitioners have turned to DFT and MD to calculate quantities including enthalpies of formation at 0 K and finite temperatures, specific heats, enthalpies of mixing, defect structures, and lattice site preferences [73, 74, 75]. DFT- and MD-predicted properties have played a critical role in informing the third generation Scientific Group Thermodata Europe (SGTE) database of the thermodynamic properties of unary systems, especially at low or high temperatures, or where phases are metastable [76]. Examples include the low temperature specific heat of numerous elements [77, 78], and free energies near and above the melting point in aluminum [76]. For multicomponent systems, the previously mentioned properties provide thermodynamic information where experiments have not or cannot be performed. For example, DFT may be used to calculate the enthalpies of formation for special configurations called end-members in a Gibbs energy description called the compound energy formalism (CEF) [79]. Furthermore, the use of these DFT enthalpies alone can provide sufficient information to specify the exact form of the CEF expressions most appropriate to a given system [80]. While it is widely understood that DFT or MD may deviate from experiments and have uncertainties deriving from several sources, few studies examine the connections between DFT and CALPHAD uncertainty. In the remainder of this section, we describe the current state of the art in propagating DFT uncertainties through CALPHAD and suggest future strategies to estimate DFT uncertainties through CALPHAD assessment and parameter fitting.

The widespread use of atomistic simulation data in the calibration of CALPHAD models has coincided with the development of strategies to fit CALPHAD parameters with uncertainty and provide probabilistic predictions, including both Bayesian [8] and frequentist [7] approaches. In 2016, Duong *et al*., described a Bayesian framework for CALPHAD uncertainty quantification and propagation and demonstrated the approach on the uranium-niobium binary system [81]. In this work, DFT calculations are performed to estimate the ground-state formation enthalpies for the γ phase, leveraging two Green's function-based approaches in addition to semi-quasi random structures (SQS). A Gaussian likelihood function was selected, and a single variance parameter was included in the inference to capture the data uncertainty. This in effect provides a single uncertainty estimate across all data, including both DFT and experimental data. Parameter inference was performed via MCMC, and then analytically propagated to phase boundaries in the binary diagram. In 2017, Duong, *et al.,* leveraged a similar framework to characterize the pseudo-binary $Ti_2AlC$-$Cr_2AlC$ phase diagrams with uncertainty [9]. In this case, finite-temperature Gibbs energies were largely provided by SQS DFT calculations, across 27 compounds and 7 temperatures, with some constraint provided via the CALPHAD models and experimental phase stability and thermodynamic information. As with the previous study, the variance in the likelihood function was fit in the Bayesian inference with a single parameter. In contrast, a novel scheme was developed that directly propagated uncertainty in the phase stability in the multicomponent space through samples from the MCMC posterior samples. This enabled a qualitative comparison of atomistic driven CALPHAD with experiment demonstrating similar levels of uncertainty. Also in 2017, Otis, *et al.*, introduced the extensible self-optimizing phase equilibrium infrastructure (ESPEI) framework for semi-automated Bayesian CALPHAD and demonstrated MCMC parameter inference in the Al-Ni system [82]. In this framework, the CEF model selection process, including the specification of sub-lattices, site ratios, and occupancies, was performed entirely using SQS enthalpies of formation and mixing. Bayesian inference was then performed using a dataset comprising 10 synthetic datasets with variance. Although not specified, we can assume that the variance in the likelihood definition was assigned as the true values for each dataset, which is common practice in the field. In 2019, Paulson, *et al.*, described a framework for the numerical propagation of uncertainty from Bayesian CALPHAD inference through MCMC for a variety of predictions used for material design tasks [83]. As a case study, the paper demonstrates inference and uncertainty propagation for the copper-magnesium binary using the ESPEI framework. In contrast to the Otis, *et al.*, study, real atomistic and experimental datasets were employed in Paulson *et al.* with reported or estimated variances (when not available or in the case of calculated data). Consequently, these variances were assigned to the Gaussian likelihood definition as weighted by a pre-factor corresponding to the data category (e.g. specific heat/enthalpy, activity, and phase stability).

Each of the preceding studies propagate the uncertainty in atomistic data forward to the CALPHAD predictions but provide no mechanism to estimate the error contribution from each data set. A possible path forward can be found in a 2019 paper from Paulson *et al.*, wherein Bayesian inference was employed to assess and calibrate models for the thermodynamic properties of elemental hafnium and rescale the reported errors for the included data sets [84]. This Bayesian approach was additionally compared to a frequentist approach in [85]. In this approach, the reported variances corresponding to each dataset served as a first guess for the variance in the likelihood. In contrast to prior work, however, each data set was assigned a unique hyperparameter that rescaled the reported variance and was included in the Bayesian inference. Through this mechanism, it was not only possible to propagate uncertainty forward, but to estimate the Bayesian scaled uncertainties associated with each dataset. The authors suggest that this same approach might be used for multicomponent systems and systems that include atomistic data. This would be a complementary mechanism to those discussed in Sec. 1 to estimate the uncertainty in DFT results. The implementation of such a scheme would encounter several challenges, most notably that this would dramatically increase the number of parameters involved in the Bayesian inference and therefore the computational expense. A potential strategy to mitigate this obstacle would be the use of approximate inference strategies that enable high-dimensional inference such as variational inference, where the shape of the posterior is assumed and the inference problem is reduced to a simple optimization [86]. Alternately, analytical gradients of the likelihood could be leveraged to accelerate Bayesian inference through Hamiltonian Monte Carlo (HMC) [87] or the No U-Turn Sampling (NUTS) approach [88].

## 5. Uncertainty Propagation in PFM meso-scale microstructure modeling

Microstructure evolution is a critical component of meso-scale modeling in materials science. The microstructure of a material strongly affects the material's properties and performance. The phase-field method is one method to model the evolution of microstructures by seeking to model phase regions.

Phase field method makes use of field variables to describe the evolutions of phase regions in time. In modern practices, evolution equations describing the evolutions of field variables in time are often derived from the thermodynamically consistent minimization of an energy functional using variational principles. An example is the Cahn-Hilliard functional [89]

$$F(x_B, \eta_k) = \int_V \left[ f_0(x_B, \eta_k) + \frac{\epsilon}{2}(\vec{\nabla} x_B)^2 + \sum_k \frac{\kappa_k}{2}(\vec{\nabla} \eta_k)^2 \right] d\vec{r} \qquad (2)$$

where $x_B$ is the concentration of phase B, $\eta_k$ is order parameter, $f_0(x_B, \eta_k)$ denotes the classical free-energy density of a homogeneous system or driving force, and the last two gradient terms represent surface tension with $\epsilon$ and $\kappa_k$ being related to interfacial energy and thickness, respectively.

As there are different energy functionals [89, 90, 91, 92], there are various phase-field models, even for the same purposes. Besides their own choices of field variables, each model features a different set of physical and/or model parameters. e.g. $\epsilon$ and $\kappa_k$ in eq. (2). Of these parameters, some can be derived from atomistic simulations with epistemic errors while others are assessed by trial-and-error approaches. Selection of models and variations of model parameters could strongly affect modeled microstructural evolutions. In this subsection, we discuss possible ways in which the uncertainties of atomistic simulation (DFT and MD-derived) derived parameters and CALPHAD derived parameters impact PFM simulated microstructure evolutions.

The most used parameters for the phase field method, derived from atomistic simulations, are interfacial energy, lattice parameter, elastic tensor, diffusion potential, and diffusion coefficient. Depending on how these parameters are conveyed to and throughout phase-field simulations, their uncertainties impact simulated microstructural evolutions differently. Figure 2 features three possible flows of physical parameters (and their uncertainties) from atomistic simulations to and throughout PFM: the first two are cross-scale, while the third is cross-time. The first cross-scale propagation is the vertical link between DFT/MD and PFM and the second is the indirect (cross link between CALPHAD and PFM) contributions of atomistic data uncertainties to CALPHAD and then to PFM. The third flow is cross-time (indicated by the differential equation within the PFM box) and is the propagation of uncertainty through PFM simulation time.

The interfacial energy, lattice parameter, elastic tensor [93], and vacancy formation energy [94] are often sourced from atomistic simulations or experiments. Correspondingly, their uncertainties are directly conveyed to PFM and their impact on the simulated microstructure evolutions is straightforward. Although diffusion potential and diffusion coefficient can be derived directly from atomistic simulation, the process can be expensive and/or is not preferred. Alternative practical approaches rely, for instance, on the use of parametric model such as Landau energy formulation to describe the thermodynamic driving force (diffusion potential) of the evolutionary system (e.g. [95, 96]) or the CALPHAD method which can be used to model both chemical potential and diffusion coefficient (e.g. [97]). For reliable thermodynamic and kinetic descriptions, these parametric approaches often adopt atomistic simulation data. This way, atomistic simulation data and their uncertainties are not conveyed directly to phase-field simulations but still contribute meaningfully to the simulations via the parametric models and their propagated uncertainties. For simplicity, the propagated uncertainties of parametric models can be seen as a composite of uncertainties coming from the atomistic simulation data, other experimental sources (if available), and the model uncertainty (i.e. the uncertainty of the model itself). It should be noted that such an uncertainty composite is not necessarily larger than the uncertainty of the atomistic simulation data. In fact, given sufficient and reliable data from various sources, it is possible that the uncertainty of the physical parameter derived from a parametric model (e.g. CALPHAD) is smaller than that calculated from DFT/MD.

In the context of indirect uncertainty propagation from atomistic simulations (like the cross link from CALPHAD shown in Figure 2), some recent notable works include Attari *et al.* [93] and Moraes *et al.* [98]. The former quantified the uncertainties of microelastic and kinetic parameters, whose ranges are biased by expert knowledge, as well as the propagation of uncertainty from CALPHAD thermodynamic driving force through the Cahn-Hilliard model. The uncertainty quantification and propagation were reliably realized by brute-force Markov chain Monte Carlo. The latter uses Landau energy model instead of CALPHAD and introduced the novel use of Probabilistic Collocation Method (a surrogate approach) integrated with sensitivity analyses to effectively reduce the computational cost required by their Monte Carlo sampling.

Whether uncertainties are passed directly or indirectly to PFM, they have to subsequently propagate through the PFM. Since a phase-field simulation is an evolutionary process, propagation of parameter uncertainties through PFM can be time-related. If a simulated evolution allows the microstructure growth to reach a steady state, the uncertainties of PFM parameters could affect the microstructure growth at the early state of the evolution but should eventually converge to the steady state. And as such, their impact can be considered time independent. If a simulated evolution was not allowed to the steady state (e.g. rapid solidification in additive manufacturing), the impact of parameter uncertainties on simulated microstructure growth through PFM could be time dependent. Often, in such a case, model parameters are functions of time. Correspondingly, their uncertainties could also evolve with time and impact the microstructure evolution in a rather complicated manner. Karayagiz, *et al.* [97], for instance, coupled a time-dependent thermal model with their phase-field model to simulate rapid solidification processes during laser powder bed fusion (L-PBF). Since Karayagiz, *et al.*, [97] adopted temperature-dependent CALPHAD chemical potential to describe their phase-field model's thermodynamic driving force, the changing temperature affects the chemical potential with time, leading to a variation of dendritic microstructures ranging from cellular to planar. Intuitively, the propagation of parameter uncertainties in time would result in magnified uncertainties of output microstructural evolutions.

**6. Challenges and outlook**

The success of multiscale modeling efforts depends on the accuracy of the individual modeling components, which for alloy design efforts frequently include PFM and CALPHAD. The inputs to these models are sometimes expensive or impossible to obtain through experimental means. This has driven the use of atomistic simulation methods, such as DFT and MD, to fill gaps in the available data. A review of the literature has revealed general rules of thumb for the expected accuracy from atomistic simulation methods. Purely DFT approaches have been shown to predict errors of up to 5 J mol$^{-1}$ K$^{-1}$ in the heat capacity of solid aluminum. MD simulations using traditional interatomic potentials can yield smaller errors of up to 1 J mol$^{-1}$ K$^{-1}$ for the same property, though care must be taken in choosing the interatomic potentials. Furthermore, *ab initio* molecular dynamics (AIMD) has shown uncertainties up to 20 K in the melting point. Although various first-principles [32, 33, 34], and CALPHAD thermodynamic and diffusion databases [99] exist that can be readily used for phase-field simulations, correspondent uncertainty databases required for UQ/UP cross-scale through PFM are not available. We believe the generation of databases that report uncertainties along with predictions will give a more confident outlook for the usage of calculated property data in PFM models.

In practice, DFT calculations of thermodynamic properties with uncertainty can be expensive for alloys and so have not been routine. While the cost is much lower for MD simulations, it is often constrained by the availability of MD interatomic potentials. Uncertainty estimates so far have been made by analyzing the effects of input parameter choices in DFT and MD simulations, using both descriptive and inferential statistics. Bayesian error estimation frameworks, which generate an ensemble of predictions, is one cost-effective path forward to calculate properties with uncertainty in DFT and MD. We believe software that automates the estimation of uncertainty, during DFT and MD simulation runs, will aid in the generation of databases of properties with uncertainty that would be helpful to CALPHAD and PFM models. For MD, several frameworks [59, 60, 70] exist that could be used to generate databases with uncertainty estimates on thermodynamic properties, especially for metals and their alloys. These databases should include calculation details and scripts to enhance reproducibility and allow users of the data to assess the methods, approximations, and limits of applicability. This approach would be facilitated by readily usable software packages that can be documented and cited; the DAKOTA framework [100] could be one step in this direction. Such calculation frameworks would also enhance assessments and comparisons of UQ methods as applied to MD simulations.

While CALPHAD modeling extensively utilizes atomistic data and uncertainty quantification has been a research topic of recent interest, no known studies examine the relationship between the uncertainty in atomistic data and in CALHPAD models. Uncertainty estimates for atomistic data would be a significant contribution to CALPHAD assessment as this could serve as weights in deterministic fitting routines or could specify the likelihood in Bayesian ones. Alternately, recently developed automated data weighting schemes could provide uncertainty estimates for atomistic data where this information is missing, either on a per-dataset basis or on the basis of another data grouping strategy (e.g. by MD interatomic potential selection). Such an approach would coincide with increased computational expense, requiring the use of more efficient inference techniques. Currently, we expect that the DFT and MD errors in the total energy would translate to a similar value for the Gibbs free energy, which affects both CALPHAD and PFM calculations. For the case of CALPHAD, this may result in up to 50 K temperature and 5 % concentration uncertainty of phase diagram features. The impact is system specific, as the shapes of the free energy curves (surfaces) play an important role in determining the accuracy and precision of the calculations.

One of the biggest challenges in uncertainty quantification and propagation of PFM is the choice of PFM model and the computational cost due to the choice of numerical solver for the model. Sometimes different phase-field models exist that share the same purposes and the same set of DFT-based/MD-based physical parameters but are different in other model parameters [101, 102]. For such models, the same input set of DFT-based/MD-based physical parameters and uncertainties would likely yield different output microstructural evolutions. Although the handling of parametric and model uncertainty coexistences exists and their propagated impacts have been studied [103], how model uncertainty affects microstructural evolution by itself and in combination with parametric uncertainty is still an open topic in PFM.

On the choice of the numerical solver, in most cases, the implementation of the phase field model makes use of the computationally expensive finite difference solver. Some models additionally require a small grid, large spatial domain size, and/or a 3-D model, thus increasing computational cost. Approaches to the propagation of DFT/MD-based parameters' uncertainties to phase-field simulation include brute-force MC that involve many such phase-field simulations sampling the parameter space, or more expensive inferential variance approaches requiring fewer simulations [88], to properly capture the effect of uncertain input parameters on output microstructural evolutions. In a few exceptional cases, an implicit or semi-implicit solver [104] can be used to accelerate a simulation while maintaining numerical stability. Consequently, studies in uncertainty quantification and propagation are generally scarce and often for cases where implicit/semi-implicit solver can be exploited [93, 98]. One approach to tackle computational cost is the use of surrogate models that improve computational efficiency for the expansive parameter sampling required for uncertainty quantification. However, special attention must be paid to preserving the key physics of the phenomenon. Otherwise, significant information loss may occur. Uncertainty quantification and propagation coupled with smart sampling of the parameter space form a reasonable methodology for evaluating the success of surrogate models. More studies are needed to understand how the uncertainties of surrogate models interplay with the uncertainties of the phase-field model's parameters [105].

The current frameworks for thermodynamic model development using DFT, MD, CALPHAD and PFM reveal that uncertainty quantification approaches exist in DFT and MD that could provide data with uncertainty to CALPHAD and PFM, but their widespread usage is limited by computational cost. For DFT and MD, we expect Bayesian error estimation frameworks to mitigate part of this cost. In the context of CALPHAD, Bayesian approaches, have gained in popularity and are facilitating robust connections that have historically been difficult to achieve. Surrogate modeling continues to be developed for accelerating uncertainty propagation studies, while minimizing accuracy loss in PFM. These developments, with increases in computational capabilities, are exciting for future simulation reliability and suggesting probable ranges of phase stability, instead of deterministic points of stability.

**Conflict of Interest**

The authors declare no competing financial interests in the writing of this manuscript.

**Acknowledgements**

J.J.G, N.H.P, and M.S. gratefully acknowledge financial support from awards 70NANB14H012 and 70NANB19H005 from U.S. Department of Commerce, National Institute of Standards and Technology as part of the Center for Hierarchical Materials Design (CHiMaD) in the Northwestern-Argonne Institute of Science and Engineering, and the Laboratory Directed Research and Development (LDRD) funding from Argonne National Laboratory, provided by the Director, Office of Science, of the U.S. Department of Energy under Contract No. DE-AC02-06CH11357. T.C.D and S.C thank the ARPA-E for its support under contract number PRJ1007310.

**Bibliography**

[1] R. Darolia, *JOM*, 43 (3), 44 (1991)

[2] G. Hanko, H. Antrekowitsch and P. Ebner, *JOM*, 54 (2), 51 (2002)

[3] B. O. Iddins, D. E. Graham, M. H. Waugh, T. Robbins, J. Cunningham III and M. T. Finn, *J. Occu. Environ. Med.*, 62 (6), 287 (2020)

[4] P. Honarmandi and R. Arróyave, *Integr. Mater. Manuf. Innov.*, 9, 103 (2020)

[5] A. V. Chernatynskiy, S. R. Phillpot and R. A. LeSar, *Ann. Rev. Mater. Res.*, 43 (1), 157 (2013)

[6] Y. Wang and D. McDowell, Uncertainty Quantification in Multiscale Materials Modeling, (San Diego: Elsevier Science and Technology), (2020)

[7] D. V. Malakhov, *Calphad*, 21 (3), 391 (1997)

[8] E. Königsberger and G. Eriksson, *Calphad*, 19 (2), 207 (1995)

[9] T. C. Duong, A. Talapatra, W. Son, M. Radovic and R. Arróyave, *Sci. Rep.*, 7, 5138 (2017)

[10] R. G. Hennig, A. Wadehra, K. P. Driver, W. D. Parker, C. J. Umrigar and J. W. Wilkins, *Phys. Rev. B.*, 82 (1), 014101 (2010)

[11] M. Stan, *Mater. Today*, 12 (11), 20 (2009)

[12] W. Kohn, A. D. Becke and R. G. Parr, *J. Phys. Chem*, 100 (31), 12974 (1996)

[13] M. P. Allen and D. J. Tildesley, Computer simulation of liquids (Oxford, UK: Clarendon Press), (1987)

[14] B. S. D. Frenkel, Understanding Molecular Simulation: From Algorithms to Applications, 2nd edition, (San Diego: Academic Press), (2002)

[15] D.C. Rapaport, The Art of Molecular Dynamics Simulation, 2nd edition, (New York: Cambridge University Press), (2004)

[16] J. Haile, Molecular Dynamics Simulation: Elementary Methods, New York: (New York, NY: Wiley–Interscience), (1997)

[17] J. Hoyt, M. Asta and A. Karma, *Mater. Sci. Eng. R*, 41 (6), 121 (2003)

[18] J. Sun, R. Haunschild, B. Xiao, I. W. Bulik, G. E. Scuseria and J. P. Perdew, *J. Chem. Phys.*, 138 (4), 044113 (2013)

[19] J. Sun, A. Ruzsinszky and J. P. Perdew, Phys. Rev. Lett., 115 (3), 036402 (2015)

[20] J. Wellendorff, K. T. Lundgaard, K. W. Jacobsen and T. Bligaard, *J. Chem. Phys.*, 140 (14), 144107 (2014)

[21] F. Tran, J. Stelzl and P. Blaha, *J. Chem. Phys.*, 144 (20), 204120 (2016)

[22] P. Janthon, S. A. Luo, S. M. Kozlov, F. Viñes, J. Limtrakul, D. G. Truhlar and F. Illas, J. Chem. Theory Comput., 10 (9), 3832 (2014)

[23] K. Choudhary, G. Cheon, E. Reed and F. Tavazza, *Phys. Rev. B.*, 98 (1), 014107 (2018)

[24] P. E. Blöchl, *Phys. Rev. B.*, 50 (24), 17953 (1994)

[25] J. J. Mortensen, L. B. Hansen and K. W. Jacobsen, *Phys. Rev. B.*, 71 (3), 035109 (2005)

[26] G. Kresse and J. Furthmüller, *Phys. Rev. B*, 54 (16), 11169 (1996)

[27] K. Lejaeghere, G. Bihlmayer, T. Björkman, P. Blaha, S. Blügel, V. Blum, D. Caliste, I. E. Castelli, S. J. Clark, A. D. Corso, S. d. Gironcoli, T. Deutsch, J. K. Dewhurst, I. D. Marco, C. Draxl, M. Dulak, O. Eriksson, J. A. Flores-Livas, Garrity, L. Genovese, P. Giannozzi, M. Giantomassi, S. Goedecker, X. Gonze, O. Grånäs, E. K. U. Gross, A. Gulans, F. Gygi, D. R. Hamann, P. J. Hasnip, N. A. W. Holzwarth, D. Iuşan, D. B. Jochym, F. Jollet, D. Jones, G. Kresse, K. Koepernik, E. Küçükbenli, Y. O. Kvashnin, I. L. M. Locht, S. Lubeck, M. Marsman, N. Marzari, U. Nitzsche, L. Nordström, T. Ozaki, L. Paulatto, C. J. Pickard, W. Poelmans, M. I. J. Probert, K. Refson, M. Richter, G.-M. Rignanese, S. Saha, M. Scheffler, M. Schlipf, K. Schwarz, S. Sharma, F. Tavazza, P. Thunström, A. Tkatchenko, M. Torrent, D. Vanderbilt, M. J. v. Setten, V. V. Speybroeck, J. M. Wills, J. R. Yates, G.-X. Zhang and S. Cottenier, *Science*, 351 (6280), 1415 (2016)

[28] K. Choudhary and F. Tavazza, *Comput. Mater. Sci.*, 161, 300 (2019)

[29] J. J. Gabriel, F. Y. C. Congo, A. Sinnott, K. Mathew, T. C. Allison, F. Tavazza and R. G. Hennig, *arXiv preprint arXiv:2001.01851*, (2020)

[30] N. L. Anderson, R. P. Vedula and A. Strachan, *Comput. Mater. Sci.*, 109, 124 (2015)

[31] J. Perdew, K. Burke and M. Ernzerhof, *Phys. Rev. Lett.*, 77 (18), 3865 (1996)

[32] A. Jain, S. P. Ong, G. Hautier, W. Chen, W. D. Richards, S. Dacek, S. Cholia, D. Gunter, D. Skinner, G. Ceder and K. A. Persson, *APL Mater.*, 1 (1), 011002 (2013)

[33] S. Curtarolo, W. Setyawan, S. Wang, J. Xue, K. Yang, R. H. Taylor, G. L. Hart, S. Sanvito, M. B. Nardelli, N. Mingo and O. Levy, *Comput. Mat. Sci.*, 58, 227 (2012)

[34] J. E. Saal, S. Kirklin, M. Aykol, B. Meredig and C. Wolverton, *JOM*, 65 (11), 1501 (2013)

[35] K. Choudhary, I. Kalish, R. Beams and F. Tavazza, *Sci. Rep.*, 7, 5179 (2017)

[36] K. Lejaeghere, V. V. Speybroeck, G. V. Oost and S. Cottenier, *Crit. Rev. Solid State Mater. Sci.*, 39 (1), 1 (2014)

[37] K. Lejaeghere, J. Jaeken, V. V. Speybroeck and S. Cottenier, *Phys. Rev. B.*, 89 (1), 014304 (2014)

[38] P.-W. Guan, G. Houchins and V. Viswanathan, *J. Chem. Phys.*, 151 (24), 244702 (2019)

[39] G. A. D. Wijs, G. Kresse and M. J. Gillan, *Phys. Rev. B.*, 57 (14), 8223 (1998)

[40] H. J. Monkhorst and J. D. Pack, *Phys. Rev. B.*, 13 (12), 5188 (1976)

[41] G. Petretto, S. Dwaraknath, H. P. C. Miranda, D. Winston, M. Giantomassi, M. V. Setten, X. Gonze, K. A. Persson, G. Hautier and G.-M. Rignanese, *Sci. Data*, 5 (1), 180065 (2018)

[42] R. Tran, Z. Xu, B. Radhakrishnan, D. Winston, W. Sun, K. A. Persson and S. P. Ong, *Sci. Data*, 3 (1), 160080 (2016)

[43] M. I. Mendelev, M. J. Kramer, C. A. Becker and M. Asta, *Phil. Mag.*, 88 (12), 1723 (2008)

[44] C. A. Becker and M. J. Kramer, *Modell. Sim. Mater. Sci. Eng.*, 18 (7), 74001 (2010)

[45] B. Grabowski, T. Hickel and J. Neugebauer, *Phys. Rev. B.*, 76 (2), 24309 (2007)

[46] C. A. Becker, F. M. Tavazza, Z. T. Trautt and R. A. B. de Macedo, *Curr. Opin. Solid State Mater. Sci.*, 17 (6), 277 (2013)

[47] L. Alzate-Vargas, M. E. Fortunato, B. Haley, C. Li, C. M. Colina and A. Strachan, *Modell. Simul. Mater. Sci. Eng.*, 26 (6), 65007 (2018)

[48] J. Mullins, Y. Ling, S. Mahadevan, L. Sun and A. Strachan, *Reliab. Eng. Syst. Saf.*, 147, 49 (2016)

[49] Z. T. Trautt, F. Tavazza and C. A. Becker, *Modell. Sim. Mater. Sci. Eng.*, 23 (7) 74009 (2015)

[50] A. P. Bartok, M. C. Payne, R. Kondor and G. Csanyi, *Phys. Rev. Lett.*, 104 (13), 136403 (2010)

[51] R. Jinnouchi, J. Lahnsteiner, F. Karsai, G. Kresse and M. Bokdam, *Phys. Rev. Lett.*, 122 (22), 225701 (2019)

[52] M. Vohra, A. Y. Nobakht, S. Shin and S. Mahadevan, *Int. J. Heat Mass Transf.*, 127, 297 (2018)

[53] R. A. Messerly, M. R. Shirts and Kazakov, *J. Chem. Phys.*, 149 (11), 114109 (2018)

[54] S. L. Frederiksen, K. W. Jacobsen, K. S. Brown and J. P. Sethna, *Phys. Rev. Lett.*, 93 (16), 165501 (2004)

[55] A. Mishra, S. Hong, P. Rajak, C. Sheng, K. Nomura, R. K. Kalia, A. Nakano and P. Vashishta, *npj Comput. Mater.*, 4 (1), 42 (2018)

[56] F. Rizzi, H. N. Najm, B. J. Debusschere, K. Sargsyan, M. Salloum, H. Adalsteinsson and O. M. Knio, *Multiscale Model. Simul.*, 10 (4), 1428 (2012)

[57] P. Zhang and D. R. Trinkle, *Modell. Simul. Mater. Sci. Eng.*, 23 (6), 65011 (2015)

[58] P. Angelikopoulos, C. Papadimitriou and P. Koumoutsakos, *J. Chem. Phys.*, 137 (14), 144103 (2012)

[59] S. Longbottom and P. Brommer, *Modell. Simul. Mater. Sci. Eng.*, 27 (4), 44001 (2019)

[60] S. T. Reeve and A. Strachan, *J. Comput. Phys.*, 334, 207 (2017)

[61] F. Rizzi, H. N. Najm, B. J. Debusschere, K. Sargsyan, M. Salloum, H. Adalsteinsson and O. M. Knio, *Multiscale Model. Sim.*, 10 (4), 1460 (2012)

[62] J. Wang, S. Olsson, C. Wehmeyer, A. Pérez, N. E. Charron, G. d. Fabritiis, F. Noé and C. Clementi, *ACS Cent. Sci.*, 5 (5), 755 (2019)

[63] F. Grogan, M. Holst, L. Lindblom and R. Amaro, *J. Chem. Phys.*, 147 (23), 234106 (2017)

[64] K. L. Joshi and S. Chaudhuri, *Phys. Chem. Chem. Phys.*, 17 (28), 18790 (2015)

[65] K. Joshi and S. Chaudhuri, *Combust. Flame*, 184, 20 (2017)

[66] K. Joshi and S. Chaudhuri, *J. Phys. Chem. C*, 122 (16), 14434 (2018)

[67] K. Lee, K. Joshi, S. Chaudhuri and D. Stewart, *Combust. Flame*, 215, 352 (2020)

[68] K. Lee, K. Joshi, S. Chaudhuri and D. S. Stewart, *J. Chem. Phys.*, 144 (18), 184111 (2016)

[69] G. Dhaliwal, P. B. Nair and C. V. Singh, *Carbon*, 142, 300 (2019)

[70] A. V. Tran and Y. Wang, *Comput. Mater. Sci.*, 127, 141 (2017)

[71] D. Zhang and S. Chaudhuri, *Comput. Mater. Sci.*, 160, 222 (2019)

[72] A. Tran, D. Liu, H. Tran and Y. Wang, *Modell. Simul. Mater. Sci. Eng.*, 27, 64005 (2019)

[73] H. Lukas, S. G. Fries and B. Sundman, Computational Thermodynamics: The Calphad Method, (Oxford, UK: Cambridge University Press), (2007)

[74] M. Stan and B. Reardon, *Calphad*, 27 (3), 319 (2003)

[75] Z.-K. Liu, *J. Phase Equilib. Diffus.*, 30 (5), 517 (2009)

[76] S. Bigdeli, L.-F. Zhu, A. Glensk, B. Grabowski, B. Lindahl, T. Hickel and M. Selleby, *Calphad*, 65, 79 (2019)

[77] J. Pavlů, P. Řehák, J. Vřešťál and M. Šob, *Calphad*, 51, 161 (2015)

[78] B. Hu, S. Sridar, L. Hao and W. Xiong, *Intermetallics*, 122, 106791 (2020)

[79] M. Hillert, *J. Alloys Compd.*, 320 (2), 161 (2001)

[80] G. Cacciamani, A. T. Dinsdale, M. Palumbo and A. Pasturel, *Intermetallics*, 18 (6), 1148 (2010)

[81] T. C. Duong, R. E. Hackenberg, A. Landa, P. Honarmandi, A. Talapatra, H. M. Volz, A. M. Llobet, A. I. Smith, G. M. King, S. Bajaj, A. Ruban, L. Vitos, P. E. A. Turchi and R. Arroyave, *Calphad*, 55, 219 (2016)

[82] R. A. Otis and Z.-K. Liu, *JOM*, 69 (5), 886 (2017)

[83] N. H. Paulson, B. J. Bocklund, R. A. Otis, Z.-K. Liu and M. Stan, *Acta Mater.*, 174, 9 (2019)

[84] N. H. Paulson, E. Jennings and M. Stan, *Intl. J. Eng. Sci.*, 142, 74 (2019)

[85] N. H. Paulson, S. Zomorodpoosh, I. Roslyakova and M. Stan, *Calphad*, 68, 101728 (2020)

[86] D. M. Blei, A. Kucukelbir and J. D. McAuliffe, *J. Am. Stat. Assoc.*, 112 (518), 859 (2017)

[87] M. Hoffman and A. Gelman, *J. Mach. Learn. Res*, 15 (1), 1593 (2014)

[88] M. Girolami and B. Calderhead, *J. R Stat. Soc. B*, 73 (2), 123 (2011)

[89] J. W. Cahn and J. E. Hilliard, *J. Chem. Phys.*, 28 (2), 258 (1958)

[90] V. Landau and L. Ginzburg, *Zh. Eksp. Teor. Fiz.*, 20, 10641082 (1950)

[91] J. Gunton, M. Miguel and P. Sahni, "The dynamics of first-order phase transitions," in Phase Transitions and Critical Phenomena vol. 8, ed. C. Domb and J.L. Lebowitz, (London, UK: Academic Press), 1987, 267-466

[92] P. Hohenberg and B. Halperin, Rev. Mod. Phys., 49 (3), 435 (1977)

[93] V. Attari, P. Honarmandi, T. Duong, D. J. Sauceda, D. Allaire and R. Arroyave, *Acta Mater.*, 183, 452 (2020)

[94] N. Wang, S. Rokkam, T. Hochrainer, M. Pernice and A. El-Azab, *Comput. Mater. Sci.*, 89, 165 (2014)

[95] P. Miles, L. Leon, R. Smith and W. Oates, *Proc. SPIE 10165*, Behavior and Mechanics of Multifunctional Materials and Composites, 1016509 (2017)

[96] L. S. Leon, R. C. Smith, P. Miles and W. S. Oates, *Proc. SPIE 10596*, Behavior and Mechanics of Multifunctional Materials and Composites XII, 105960T (2018)

[97] K. Karayagiz, L. Johnson, R. Seede, V. Attari, B. Zhang, X. Huang, S. Ghosh, T. Duong, I. Karaman, A. Elwany and R. Arróyave, Acta Mater, 185, 320 (2020)

[98] E. A. B. de Moraes, M. Zayernouri and M. M. Meerschaert, *Int. J. Numer. Methods. Engg.*, vol. submitted.

[99] R. Schmid-Fetzer, D. Andersson, P.-Y. Chevalier, L. Eleno, O. Fabrichnaya, U. Kattner, B. Sundman, C. Wang, A. Watson, L. Zabdyr and M. Zinkevich, *Calphad*, 31 (1), 38 (2007)

[100] M. Wood, M. Cusentino, B. Wirth and A. Thompson, *Phys. Rev. B*, 99, 184305 (2019)

[101] I. Steinbach, L. Zhang and M. Plapp, *Acta Mater.*, 60 (6), 2689 (2012)

[102] S. G. Kim, W. T. Kim and T. Suzuki, *Phys. Rev. E*, 60 (6), 7186 (1999)

[103] P. Honarmandi, T. Duong, Ghoreishi, D. Allaire and R. Arroyave, *Acta Mater.*, 164, 636 (2019)

[104] L. Chen and J. Shen, *Comput. Phys. Comm.*, 108 (2), 147 (1998)

[105] D. E. Ricciardi, O. A. Chkrebtii and S. R. Niezgoda, *Integr. Mater. Manuf. Innov.*, 9 (2), 181 (2020)